# Optimization of Delivery Routes for Fresh E-commerce in Pre-warehouse Mode

Alice Harward[1], Junjie Lin[1], Yun Wang[2], Xiaoke Xie[3]

1. Shanghai University of International Business and Economics No.1900 Wenxiang Road, Songjiang District, Shanghai, China 201620
2. Shanghai University of Finance and Economics - SUFE， Yangpu, Shanghai, China. 11757
3. Business School, Shanghai Industry & Commerce Foreign Language College, New Pudong District, Shanghai, China, 201210

**Abstract**

With the development of the economy, fresh food e-commerce has experienced rapid growth. One of the core competitive advantages of fresh food e-commerce platforms lies in selecting an appropriate logistics distribution model. This study focuses on the front warehouse model, aiming to minimize distribution costs. Considering the perishable nature and short shelf life of fresh food, a distribution route optimization model is constructed, and the saving mileage method is designed to determine the optimal distribution scheme. The results indicate that under certain conditions, different distribution schemes significantly impact the performance of fresh food e-commerce platforms. Based on a review of domestic and international research, this paper takes Dingdong Maicai as an example to systematically introduce the basic concepts of distribution route optimization in fresh food e-commerce platforms under the front warehouse model, analyze the advantages of logistics distribution, and thoroughly examine the importance of distribution routes for fresh products.

**Keywords:** front warehouse, fresh food e-commerce, distribution route optimization

# 1 Introduction

## 1.1 Research Background

With the advancement of the times and the progress of internet technology, China's fresh food e-commerce industry has developed rapidly. The emergence of the front warehouse model has gained widespread popularity due to its low construction costs, high delivery efficiency, and simplified operational processes. However, the low entry barriers and intense competition in the fresh food e-commerce industry, coupled with the inherent limitations of the front warehouse model, have posed challenges to improving delivery efficiency. Against the backdrop of customer demand, selecting the optimal delivery model and route to maximize corporate profits has become an urgent issue, with transport route optimization being the most critical factor.

Fresh food plays a vital role in enhancing people's quality of life. In China, in particular, dietary habits have shifted from being grain-based to fruit-and-vegetable-focused, driving the rapid growth of the fresh food e-commerce sector and creating vast opportunities for development in the fresh food logistics field. Currently, China's fresh food e-commerce industry faces challenges such as high logistics costs and low efficiency in logistics management. In response, major fresh food e-commerce platforms have been actively exploring innovative logistics methods, among which the "front warehouse" model stands out as the most representative. By studying delivery route optimization for Dingdong Maicai under the front warehouse model, this paper aims to establish an efficient logistics distribution system for China's fresh food e-commerce platforms and provide logistics optimization strategies for similar enterprises.

## 1.2 Research Significance

Studying path optimization strategies for fresh food e-commerce platforms under the front warehouse model holds significant practical and academic value. First, the front warehouse model enhances operational efficiency and ensures product freshness through centralized procurement, storage, and distribution. However, with increasing market competition and evolving consumer demands, optimizing delivery routes under the front warehouse model to reduce costs and improve efficiency has become crucial. This optimization not only involves the routes between front warehouses but also the routes from logistics centers to front warehouses. Therefore, research on route optimization at both levels carries substantial practical importance for fresh food e-commerce platforms.

Second, Dingdong Maicai has demonstrated remarkable success in applying the front warehouse model, offering valuable lessons for other fresh food e-commerce platforms. An in-depth analysis of Dingdong Maicai’s path optimization can not only summarize its successful practices but also identify existing issues, providing guidance for future improvements.

Moreover, optimizing delivery routes contributes to the overall development of the fresh food e-commerce industry. By reducing operational costs and enhancing delivery efficiency, platforms can increase customer satisfaction and strengthen their competitiveness, fostering the industry’s prosperity.

Finally, from an academic perspective, this research enriches the theoretical framework of logistics in the fresh food e-commerce field. By exploring route optimization methods for both logistics centers to front warehouses and between front warehouses, this study provides new insights and directions for future research.

# 2 Literature Review

In recent years, research on fresh food e-commerce based on the front warehouse model has developed rapidly. Current studies mainly focus on two aspects: development status and location optimization. Regarding development status, Wang Chengrong (2020) [1] analyzed the operational characteristics of the front warehouse model in fresh food e-commerce from a supply chain perspective, providing suggestions and solutions for its long-term development. In terms of location and optimization, researchers have analyzed the advantages and disadvantages, using methods like big data analysis for optimization. Jiao Zhenbo (2020) [2], for instance, constructed a model for front warehouse locations and layouts that minimize total logistics costs under a "regional distribution center + front warehouse" layout model. Du et.al. [3] developed two operator models in which the front warehouse's inventory services facilitate vehicle capacity recycling, reducing the number of vehicles used and lowering logistics and distribution costs. Wang Yan (2023) [4] interpreted front warehouses from a supply chain perspective, analyzing their contribution to supply chain development and proposing development strategies based on the supply chain.

X.Du et al (2022) [5] used Missfresh as a case study to summarize and analyze supply chain strategies and demand forecasting in the front warehouse model, highlighting issues such as limited product variety and single profitability models in current supply chains. Huang Zuoming and Ding Yingwenxiu (2021) [6] analyzed the cold chain logistics process of Dingdong Maicai, identifying existing problems and proposing corresponding cost control measures. Fresh food e-commerce, which attracts consumer loyalty, occupies a significant position in the market; however, its short shelf life, high wastage, and storage difficulties heavily depend on cold chain logistics, making cost control crucial. Wen Zhenxin et al. (2019) [7] analyzed traditional fresh food e-commerce, noting the emergence of innovative products, with the front warehouse model as a representative. Compared to traditional fresh food e-commerce, front warehouses offer significant advantages in location and scale effects.

Xia Zixin et al. (2023) [8] studied supply chain management in the context of the SCOR model, proposing optimization measures in planning, procurement, production, distribution, and returns. The outbreak of COVID-19 and the rise of e-commerce have driven demand for solutions to the perishability and fragility of fresh products, challenges that traditional logistics struggle to address. Fan Bingpeng (2023) [9] studied fresh cold chain delivery, considering costs such as vehicle fixed costs, transportation costs, product loss costs, and time window penalty costs, constructing an optimization model for vehicle routing under time window constraints. Wang Hairuo (2022) [10] analyzed value positioning, value network, value maintenance, and value realization, addressing issues such as single profitability models and product homogenization in community fresh food business models. Strategies for business model innovation were proposed to offer insights into the development of fresh food e-commerce in China.

Zeng Xiaoke and Zeng Yaorui (2022) [11], under the impact of the pandemic, researched the supply chain delivery model of Dingdong Maicai, suggesting improvements using SOP (Standard Operating Procedures) to promote cost reduction and efficiency enhancement. This approach aims to achieve higher GMV (Gross Merchandise Value) for the platform. The "front warehouse" model, represented by platforms like Dingdong Maicai, JD Daojia, and Hema Maicai, has garnered widespread attention. Li Zhen and Wang Juan (2022) [12] used Dingdong Maicai as an example to analyze the supply chain status under the front warehouse model, highlighting its problems and offering suggestions for improvement. To address various issues in the storage and distribution processes of fresh products, many enterprises have implemented and adopted different warehouse distribution models. Fresh food, as a daily necessity for consumers, is seeing increasing maturity in its e-commerce logistics development. Scholars have focused on the following aspects:

1. Transportation Research in Fresh Food E-Commerce Logistics**: Li Guimei (2021) [13] summarized the advantages and disadvantages of various delivery models, including self-built cold chain logistics, third-party delivery, and consumer self-pickup, based on an analysis of issues in fresh food e-commerce logistics. Huang Jidan (2022) [14] developed a mixed-integer programming model considering delivery priority, costs, and time, solving it with an improved krill foraging optimization algorithm and validating its effectiveness through simulations.

2. Development Research in Fresh Food E-Commerce Logistics**: Lü Zihui (2018) [15] constructed a standard system for cold chain logistics in fresh food e-commerce based on national, regional, and industry standards and the development status in China. Xing Xianghuan (2020) [16] analyzed problems in fresh food e-commerce logistics distribution models and proposed solutions such as optimizing cold chain logistics systems and fresh product supply chains.

3. Service Quality Evaluation Research in Fresh Food E-Commerce Logistics**: Xu Guangshu (2021) [17] added product quality and value-added services to existing service quality evaluation systems, using rough set theory to determine evaluation indicator weights. Consumer perceptions and expectations were collected through surveys to evaluate logistics service quality. Li Shaoying et al. (2022) [18] developed a 22-indicator system for evaluating fresh food e-commerce logistics based on SERVQUAL and LSQ classic models, incorporating dimensions like delivery time, information services, and service standardization. The AHP method was used to determine indicator weights. Du X et al. (2022) [19] constructed a 20-indicator system for fresh food e-commerce logistics service quality evaluation using similar models and conducted empirical studies.

In modern logistics, distribution is a critical element, responsible for transporting goods to meet demand and serving as the final stage of the logistics process. Distribution route optimization aims to minimize costs, routes, and transport distances while satisfying customer service requirements. As a highly complex nonlinear programming problem, scholars worldwide have conducted extensive research on it, achieving significant progress. Rekabi S (2024) [20] developed a sustainable vehicle routing optimization model for solid waste networks from an IoT dual-objective perspective. Varas M (2024) [21] addressed distribution routes with multiple time windows, mandatory returns, and perishable items, emphasizing adaptive adjustments for time constraints and perishability requirements. Lehmann J and Winkenbach M (2024) [22] analyzed a mathematical model for a two-level multi-trip vehicle routing problem with mixed pickup-and-delivery needs under time constraints. Oliver and Smith; William Ho (2008) [23] and Hagiwara A (1993) [24] conducted comprehensive research on route optimization. Taniguchi (2001) [25] found that many companies adopt shared delivery methods to reduce logistics costs. Ballou (2012) [26] examined issues like transport mode selection, route planning, vehicle scheduling, and centralized dispatch. Dantzig and Ramser (1959) [27] proposed a freight route planning method based on the shortest route to meet station capacity needs simultaneously. Naji-AzimiZ (2016) [28] introduced a method based on shortest routes and maximum capacity. Cao (2017) [29] integrated dynamic traffic flows with user fuzzy time windows to construct a novel vehicle route model. Hernandez F (2019) [30] proposed a two-stage planning model for uncertain environments, aiming to minimize route and expected costs. Zulvia F E (2020) [31] built a green logistics route optimization model for perishable goods, considering time sensitivity, peak and off-peak traffic, operating time, and factors like operating costs, spoilage costs, carbon emissions, and customer satisfaction. Lin (2021) [32] optimized GOME's small-batch, multi-frequency distribution model, reducing overall logistics costs using a mileage-saving method.

# 3Overview of Dingdong Maicai’s Business Model

## 3.1 Introduction to Dingdong Maicai

Dingdong Maicai, established in 2017, is an online fresh food e-commerce platform focusing on delivering fresh produce such as fruits, vegetables, meat, and seafood to urban residents. Its innovative business model, based on a "front warehouse + instant delivery" strategy, aims to meet consumers' high demand for freshness and rapid delivery. The key components of its business model are as follows:

**1.Demand Forecasting**:

Using big data technology, Dingdong Maicai predicts customer demand based on historical orders, seasonal changes, and other factors. This enables the platform to prepare inventory in advance, ensuring supply chain stability and timely delivery.

**2.Direct Procurement**:

Dingdong Maicai primarily sources products from urban wholesale markets and branded suppliers. Compared to traditional procurement methods, this flexible approach shortens delivery routes, reduces cold-chain logistics costs, stabilizes product prices, and improves restocking efficiency.

**3. Intelligent Platform**:

The platform employs big data analytics and intelligent recommendation systems to enhance order conversion rates and customer satisfaction. Accurate recommendations and product optimization streamline the shopping experience.

**4.Optimized Logistics System**:

Dingdong Maicai uses big data to optimize delivery routes and integrates a self-built logistics system to achieve fast doorstep deliveries. It establishes front warehouses in communities, with each warehouse handling orders within its area. Orders are delivered within 29 minutes, and if a warehouse exceeds 1,500 daily orders, it is split into two to maintain efficiency. The "zero delivery fee + no minimum order" policy further caters to consumers' instant needs.

**5.After-Sales Service and Feedback Collection**:
Post-delivery, the platform collects customer feedback to generate data for optimizing products and services. Comprehensive after-sales services ensure continuous improvement in the customer experience.

Dingdong Maicai also operates smart cold storage facilities that enhance product sorting and management. Advanced temperature control technologies maintain product quality during storage and transportation. Its "one-hour delivery" model, supported by an efficient logistics system, improves user experience and strengthens market competitiveness. During the pandemic, Dingdong Maicai gained significant market share in Shanghai’s fresh food market, exemplifying the city’s efficiency. Its ability to meet urgent consumer demand cemented its position as a trusted fresh food e-commerce platform.

Dingdong Maicai’s "central warehouse + front warehouse" operation involves consolidating goods at central warehouses before distributing them to front warehouses via a streamlined logistics system. This model improves efficiency and responsiveness to consumer needs. The semi-open front warehouses, located near communities, double as storage facilities and mini retail spaces, offering diverse products and quick delivery options that cater to daily fresh food needs.

## 3.2 Operational Status of Dingdong Maicai

Through supply chain research, Dingdong Maicai’s goals across upstream, midstream, and downstream operations are as follows:

**Upstream**:
Focus on urban wholesale procurement, quality control, and optimizing supply chain stability to enhance competitiveness. A dedicated procurement team ensures high product quality, efficient sorting, and cost control.

**Midstream**:
Leverage big data for product management, cost control, marketing, and quality assurance. Strategies include minimizing reverse logistics costs, improving customer experience, and promoting community-based marketing methods for rapid user acquisition. The "Today's Menu" feature encourages cooking and boosts user engagement and repeat purchases.

**Downstream**:
Despite the high cost, Dingdong Maicai's self-built cold chain logistics ensures product freshness and a superior customer experience. The "zero delivery fee + no minimum order" model attracts numerous customers.

# 4 Optimization of Delivery Routes Under the Front Warehouse Model

## 4.1 Goals and Principles of Route Optimization

Transportation route selection significantly impacts costs. Based on network topology optimization, this study addresses distribution from Dingdong Maicai's logistics center to front warehouses. Route optimization principles include:

**Minimizing Distance**: Shorten total driving distance without compromising service quality.

**Reducing Costs**: Select the most cost-effective route while considering product characteristics.

**Customer Satisfaction**: Focus on timely delivery and quality to enhance competitiveness.

**Time Efficiency**: Prioritize time savings to reduce overall costs.

**Safety and Reliability**: Minimize risks during transportation to ensure product integrity.

By integrating these principles, businesses can achieve optimal logistics routes tailored to operational needs.

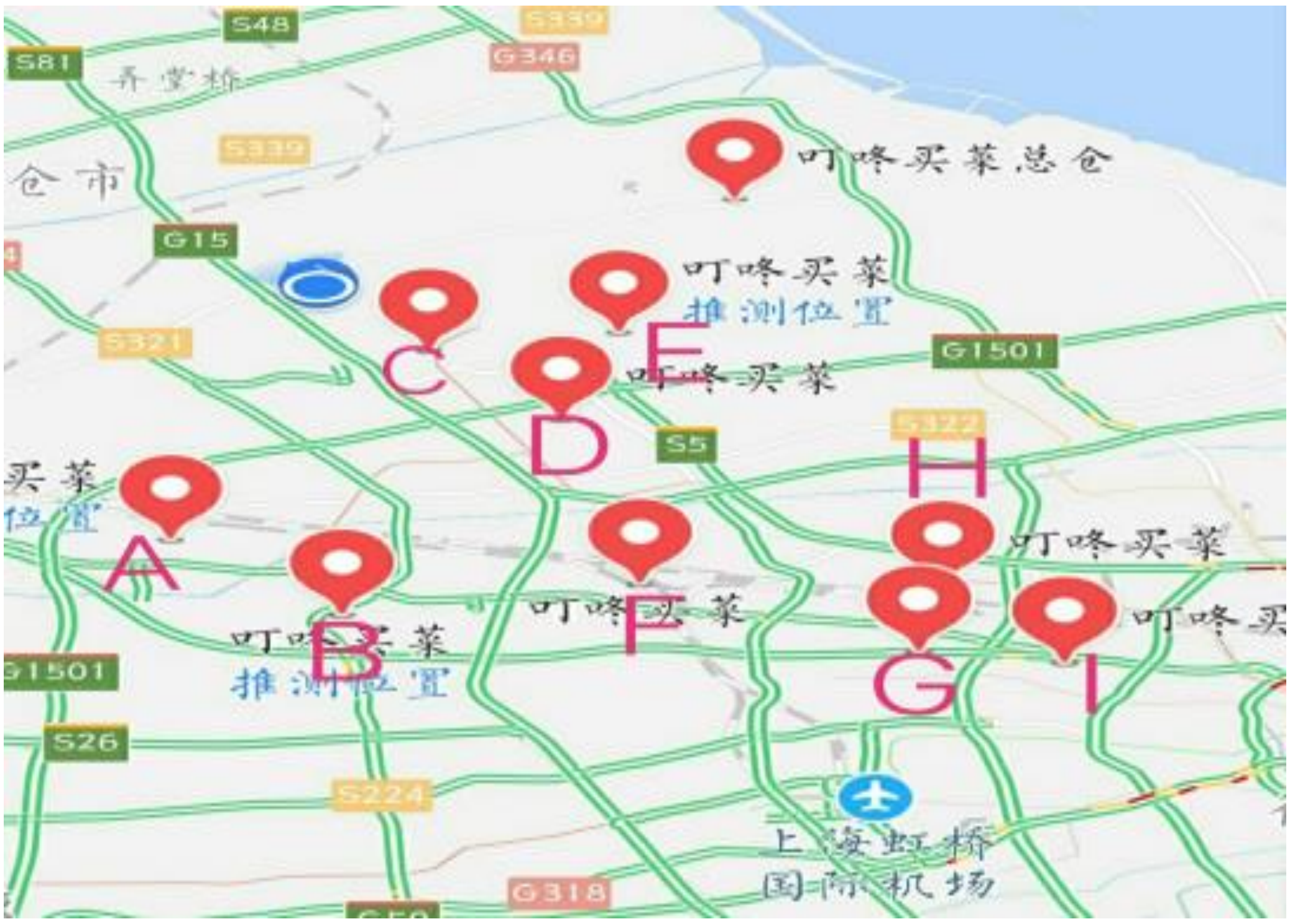

Figure 4-1: Distribution of Front-End Warehouses in a Selected Area of Shanghai

**Saved Mileage Method**: The core concept is to deliver goods to LL distribution points using NN vehicles. This method aims to minimize transportation costs by establishing the minimum number of routes MM required in the transportation network and optimizing the scheduling of the network.

Firstly, this method combines two stages of the logistics system into a single loop, enabling customers to complete goods delivery in a timely manner, thereby saving vehicle travel time and costs. Secondly, after integrating the loop, a corresponding variable value—known as the "saved mileage"—emerges due to the reduced transmission distance. When using the saved mileage method for calculations, the process can be conducted in three steps:

1. Generate an initial solution to determine the optimal allocation of demand and constraints for each front-end warehouse.

2. Calculate the savings degree ΔCij through computation.

3. Arrange the saved mileage in descending order and combine the routes accordingly.

The distances between the central warehouse and the front-end warehouses, as well as those between the front-end warehouses themselves, are as follows (in kilometers):

Table 4-1 Distances Between Central Warehouse and Front Warehouses, and Among Front Warehouses

| | P | | | | | | | | |
|---|---|---|---|---|---|---|---|---|---|
| A | 30 | A | | | | | | | |
| B | 31 | 3.2 | B | | | | | | |
| C | 14 | 9 | 10.6 | C | | | | | |
| D | 16 | 6.2 | 8.5 | 3 | D | | | | |
| E | 9.6 | 5.8 | 10 | 3 | 3 | E | | | |
| F | 24 | 6.7 | 3.8 | 11 | 7 | 8.6 | F | | |
| G | 31 | 13 | 8 | 17 | 15 | 13 | 4 | G | |
| H | 27 | 14 | 11 | 15.3 | 12 | 14 | 4 | 3 | H |
| I | 32 | 16 | 11 | 20 | 18 | 16 | 6 | 3 | 5 |

The Demand of Front Warehouses is Listed Below, with a Truck's Rated Load Capacity of 8 Tons:

Table 4-2 Demand of Front Warehouses

| | A | B | C | D | E | F | G | H | I |
|---|---|---|---|---|---|---|---|---|---|

| Demand quantity | 1.3 | 1.0 | 1.5 | 1.7 | 1.6 | 1.5 | 1.3 | 1.5 | 1.4 |
|---|---|---|---|---|---|---|---|---|---|

Table 4-3 Saved Mileage Between Front Warehouses

| | A | | | | | | | |
|---|---|---|---|---|---|---|---|---|
| B | 57.8 | B | | | | | | |
| C | 35 | 34.4 | C | | | | | |
| D | 39.8 | 38.5 | 27 | D | | | | |
| E | 33.6 | 30.6 | 22.6 | 32.6 | E | | | |
| F | 47.3 | 61.2 | 27 | 33 | 25 | F | | |
| G | 48 | 54 | 28 | 32 | 27.6 | 31 | G | |
| H | 43 | 47 | 25.7 | 31 | 22.6 | 47 | 55 | H |
| I | 46 | 52 | 26 | 30 | 25.6 | 50 | 60 | 54 |

Table 4-4 Sorting of Saved Mileage by Size

| **No.** | **Connection** | **Saved Mileage** | **No.** | **Connection** | **Saved Mileage** | **No.** | **Connection** | **Saved Mileage** |
|---|---|---|---|---|---|---|---|---|
| 1 | B-F | 61.2 | 13 | A-I | 46 | 25 | B-E | 30.6 |
| 2 | G-I | 60 | 14 | A-H | 43 | 26 | D-I | 30 |
| 3 | A-B | 57.8 | 15 | A-D | 39.8 | 27 | C-G | 28 |
| 4 | G-H | 55 | 16 | B-D | 38.5 | 28 | E-G | 27.6 |
| 5 | H-I | 54 | 17 | A-C | 35 | 29 | C-F | 27 |
| 6 | B-G | 54 | 18 | B-C | 34.4 | 30 | C-D | 27 |
| 7 | B-I | 52 | 19 | A-E | 33.6 | 31 | C-I | 26 |
| 8 | F-I | 50 | 20 | D-F | 33 | 32 | C-H | 25.7 |

| No. | Connection | Saved Mileage | No. | Connection | Saved Mileage | No. | Connection | Saved Mileage |
|---|---|---|---|---|---|---|---|---|
| 9 | A-G | 48 | 21 | D-E | 32.6 | 33 | E-I | 25.6 |
| 10 | A-F | 47.3 | 22 | D-G | 32 | 34 | E-F | 25 |
| 11 | B-H | 47 | 23 | D-H | 31 | 35 | C-E | 22.6 |
| 12 | F-H | 47 | 24 | F-G | 31 | 36 | E-H | 22.6 |

**Initial Solution**: The distribution is carried out from the central warehouse P to each of the nine forward warehouses, with a total operating distance of 214.6 km and requiring 9 trucks. The diagram is as follows:

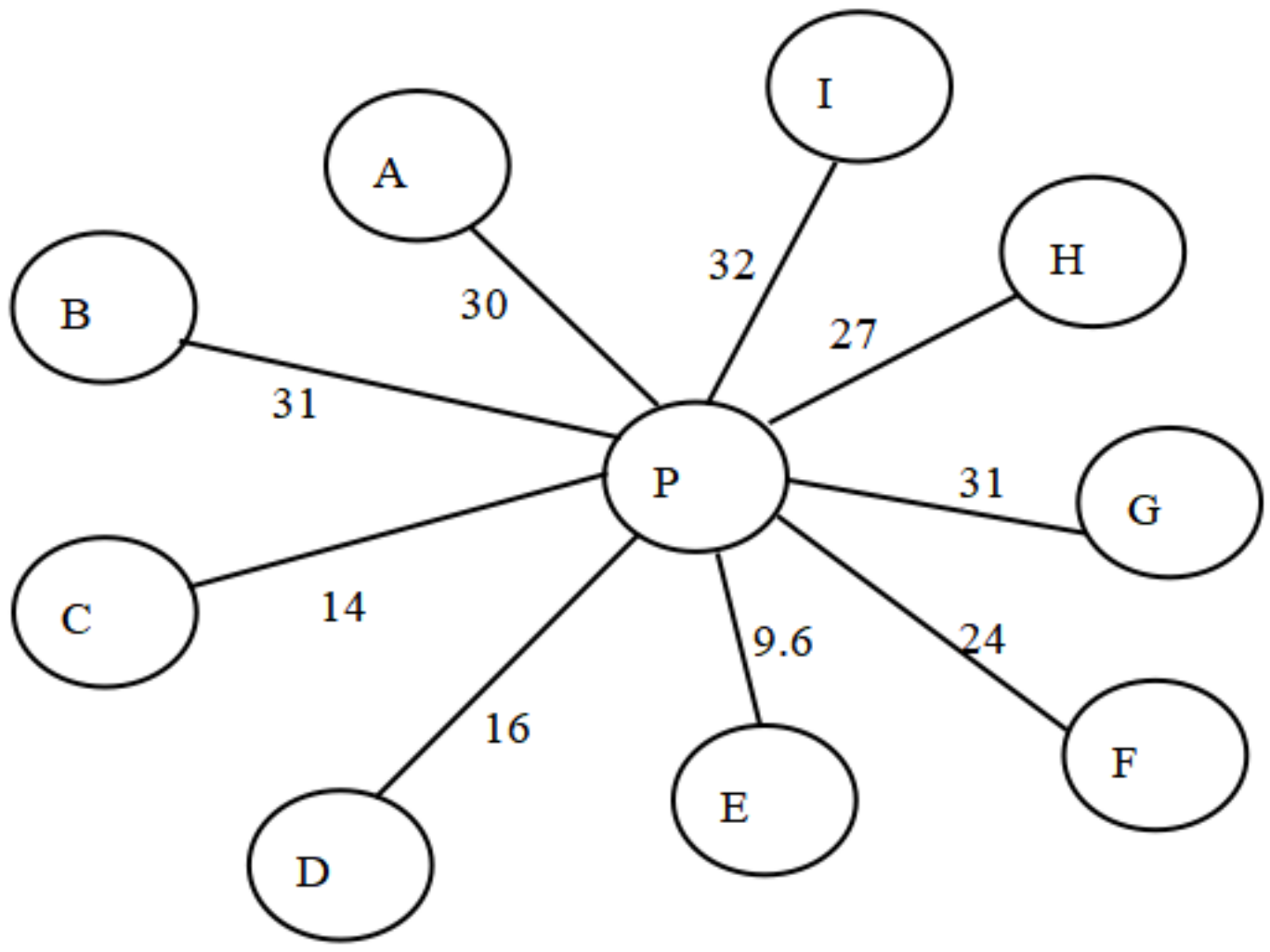

Fig. 4-2 Initial Solution

**Secondary Solution**: The routes are adjusted based on the sorted savings in mileage, connecting B-F and B-A. The distribution routes now total 7, with a total operating distance of 190.6 km and requiring 7 trucks. The diagram is as follows:

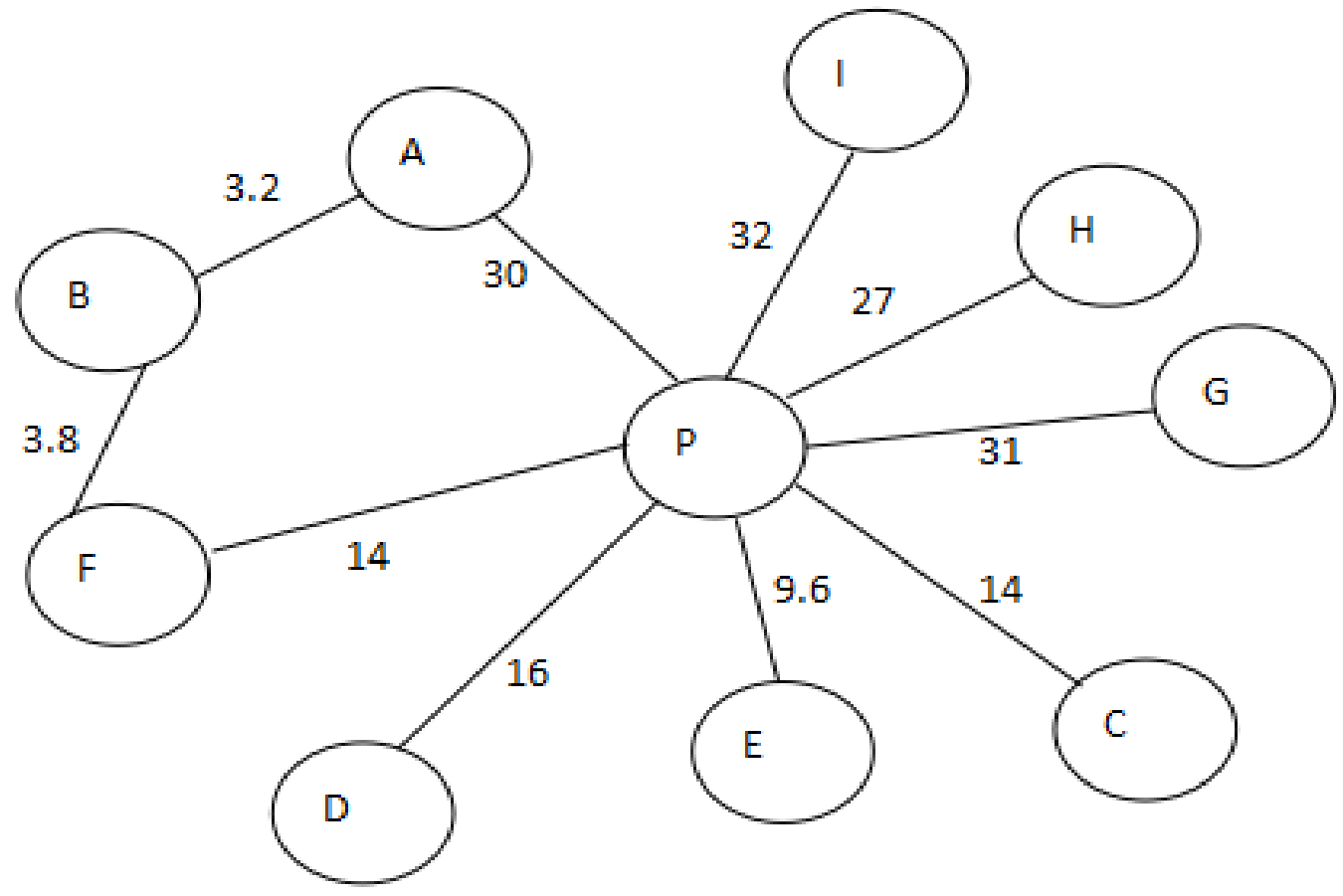

Fig.4-3 Second Solution

**Tertiary Solution:** The routes are adjusted by connecting A-G and G-I. At this point, connecting any other warehouse would exceed the maximum truck capacity. Therefore, A, B, F, G, and I form a closed loop with the central warehouse P, creating a single distribution route. The total number of routes is now 5, with a total operating distance of 146.5 km, requiring 5 trucks. The diagram is as follows:

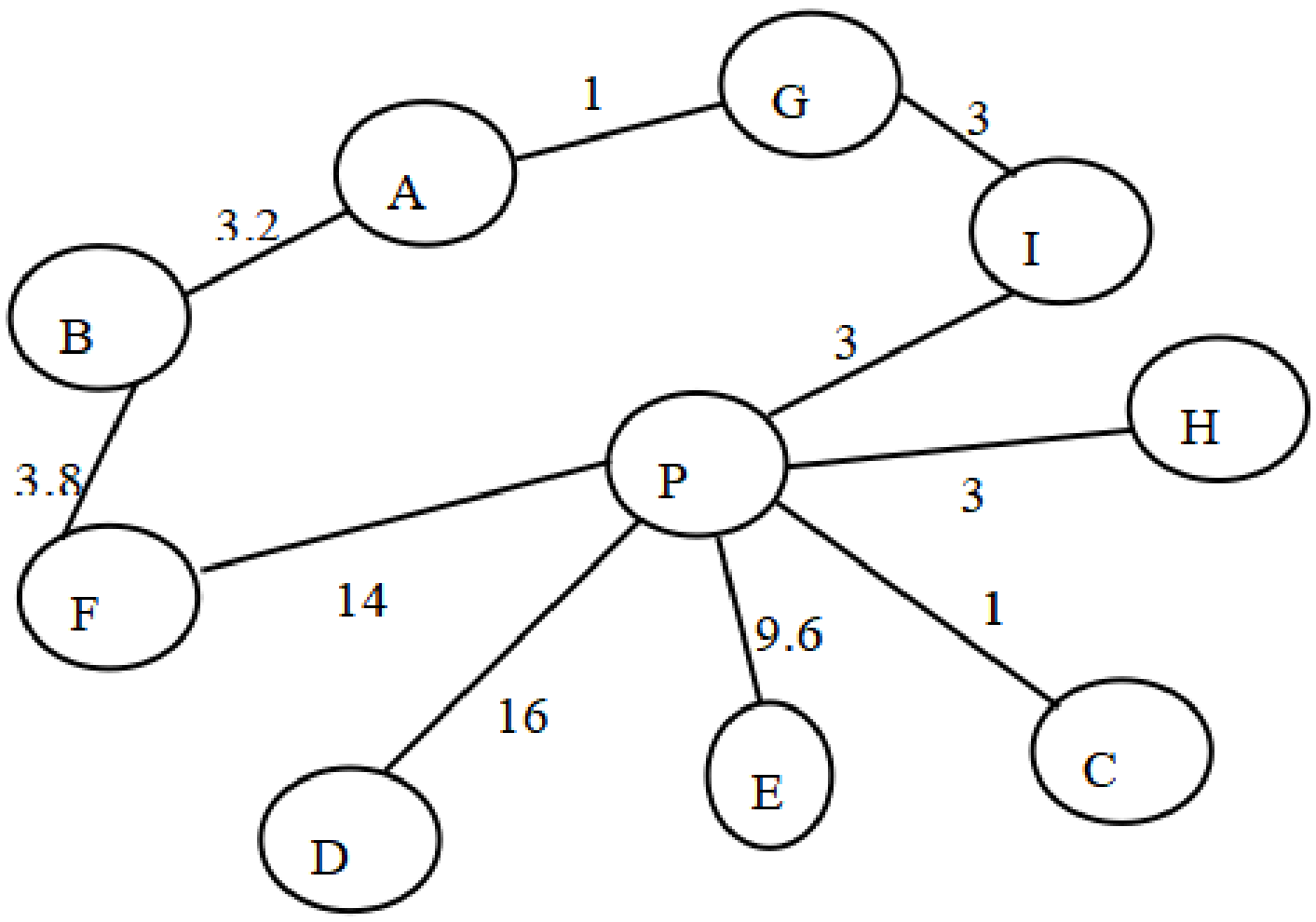

Fig. 4-4 Tertiary Solution:

**Final Solution:** After calculation, the total demand of the remaining four warehouses falls within the maximum truck capacity of 8 tons. Therefore, these four warehouses are merged into one distribution route. At this point, there are 2 distribution routes in total, with a total operating distance of 122.9 km, requiring 2 trucks. The diagram is as follows:

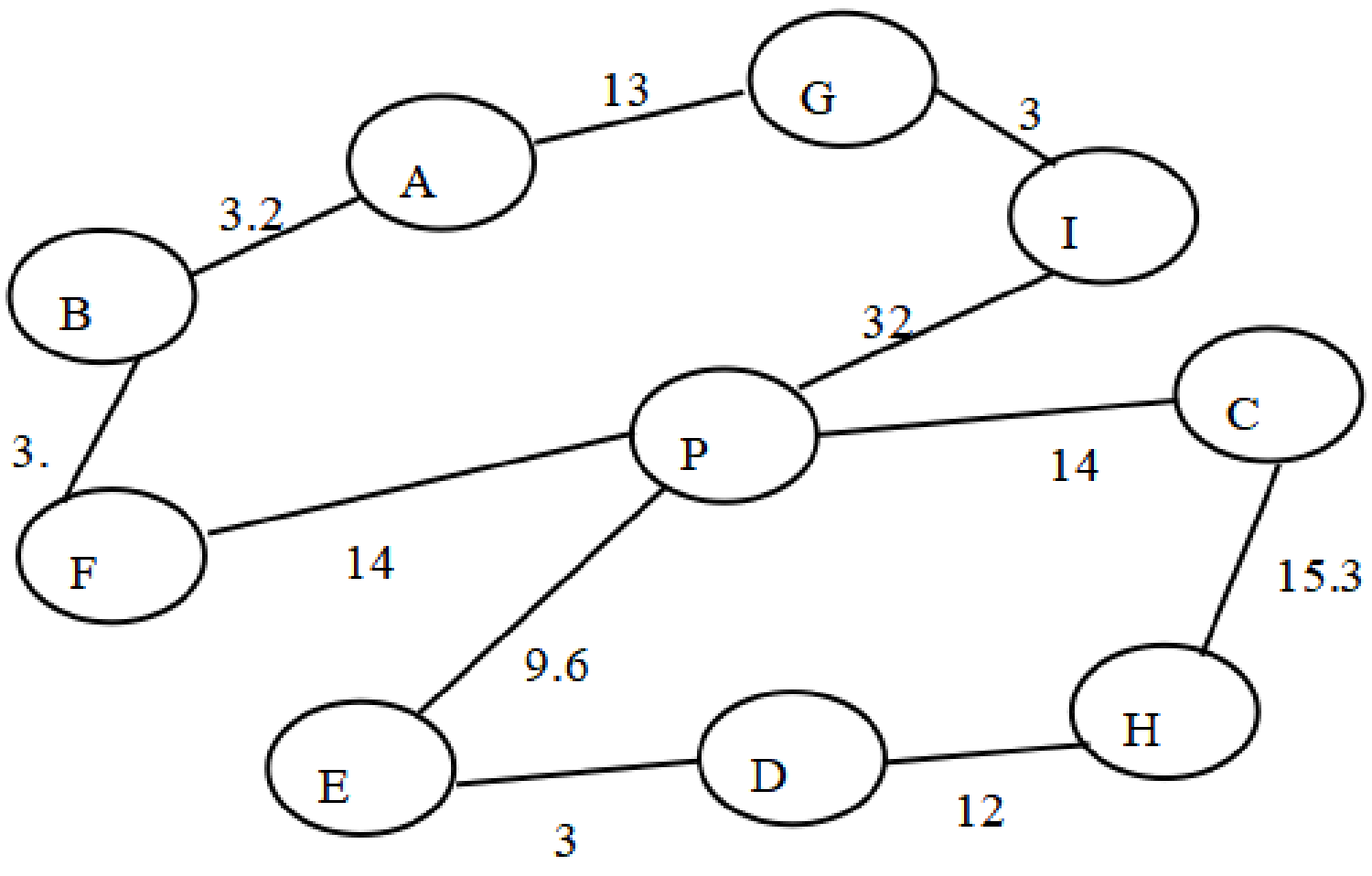

Fig.4-5 Final Solution

## 4.2 Path Optimization Summary

This study primarily focuses on ordinary goods that do not require storage and have long shelf lives. These products vary in type and are in high demand in the market, with a wide range of transportation tools available. To optimize the distribution process, this paper proposes a savings mileage method that takes market demand into account. Given the limited loading capacity of trucks, it is necessary to appropriately organize the transportation paths to maximize profit and reduce empty vehicle rates. This can improve the utilization efficiency of transportation tools and save costs.

Through calculations, the total demand of the remaining four warehouses falls within the maximum truck capacity of 8 tons. Therefore, these four warehouses are merged into a single distribution route. The optimized distribution plan includes two main routes, with a total operating distance of 122.9 kilometers, requiring two trucks. This optimization allows the two trucks to simultaneously complete their respective tasks, thus saving transportation tools and shortening transportation time, which significantly enhances work efficiency.